\documentclass[12pt,a4paper]{article}
\pdfoutput=1
\usepackage{jheppub, hyperref, amsmath, amssymb, graphicx, amsfonts, latexsym, bbold, mathbbol, color, slashed, bigints}
\newcommand{\be}{\begin{equation}}
\newcommand{\ee}{\end{equation}}
\newcommand{\bea}{\begin{eqnarray}}
\newcommand{\eea}{\end{eqnarray}}
\def\eqa{&=&} 
\def\ccr{\nonumber\\} 
\def\la{\langle}
\def\ra{\rangle}

\author[a,c]{Fiorenzo Bastianelli,}
\author[b,c]{Olindo Corradini,}
\author[b]{Laura Iacconi}

\affiliation[a] {Dipartimento di Fisica ed Astronomia, Universit{\`a} di Bologna,
via Irnerio 46, I-40126 Bologna, Italy}
\affiliation[b]  {Dipartimento di Scienze Fisiche, Informatiche e Matematiche, Universit\`a di Modena e Reggio Emilia, Via Campi 213/A, I-41125 Modena, Italy}
\affiliation[c] {INFN, Sezione di Bologna, via Irnerio 46, I-40126 Bologna, Italy}

\emailAdd{fiorenzo.bastianelli@bo.infn.it} 
\emailAdd{olindo.corradini@unimore.it}
\emailAdd{180851@studenti.unimore.it}

\abstract{Particles in a curved space are classically described by a nonlinear sigma model action that can be quantized
through path integrals. The latter require a precise regularization to deal with the derivative interactions arising from the 
nonlinear kinetic term. Recently, for maximally symmetric spaces, simplified path integrals have been developed: they allow 
to trade the nonlinear kinetic term with a purely quadratic kinetic term (linear sigma model). This happens at the expense of introducing 
a suitable effective scalar potential, which contains the information on the curvature of the space.
The simplified path integral provides a sensible gain in the efficiency of perturbative calculations.
Here we extend the construction to models with $N=1$ supersymmetry on the worldline, which are applicable to the 
first quantized description of a Dirac fermion. As an application we use the simplified worldline path integral to compute 
the type-A trace anomaly of a Dirac fermion in $d$ dimensions up to $d=16$.}

\title{Simplified path integral for supersymmetric quantum mechanics and type-A trace anomalies}

\begin{document}
\maketitle
\flushbottom

\section{Introduction}
Recently,  a simplified version of the path integral for the quantum mechanics of particles on maximally symmetric spaces
has been constructed \cite{Bastianelli:2017wsy, Bastianelli:2017xhi}. It realizes an old proposal \cite{Guven:1987en},
which suggests a peculiar use of Riemann normal coordinates to trade the nonlinear kinetic term of the classical action 
of the particle with a purely quadratic kinetic term (linear sigma model) 
at the expense of introducing a suitable effective scalar potential.  The conjecture was originally made for an arbitrary curved space,
but the explicit proof presented in \cite{Bastianelli:2017wsy, Bastianelli:2017xhi} works only for spaces with 
maximal symmetry. The more subtle question of its validity on arbitrary geometries remains open, though
a positive answer seems unlikely.

In the present paper we extend the construction to a $N=1$ supersymmetric quantum mechanics,
so that the resulting path integral can be used in a first quantized description of a Dirac fermion.
In particular,  we use the new path integral to compute the type-A trace anomalies of a Dirac field, up to $d=16$ dimensions, extending 
analogous calculations performed  in \cite{Bastianelli:2017wsy, Bastianelli:2017xhi} for the conformal scalar.

Other methods for identifying the type-A trace anomalies for the spin-1/2 field in higher dimensions are probably more efficient. 
One may use the zeta function approach as in \cite{Copeland:1985ua, Cappelli:2000fe}, 
or the AdS/CFT holographic paradigm as in~\cite{Aros:2011iz}, which extended to spin 1/2 
the scalar case treated in~\cite{Diaz:2008hy,Dowker:2010qy}.  
However, the path integral method has a clearer physical interpretation.
It is a calculation from first principles, in which the particle producing the one-loop
anomaly performs its virtual loop. This visualization makes the method more intuitive and flexible, 
allowing to study many other observables, as typical in the worldline formalism. 
The latter employs worldline path integral to represent and compute effective actions
and scattering amplitudes, see  \cite{Schubert:2001he} for a review in flat space, 
refs.~\cite{Dunne:2005sx, Dunne:2006st, Ahmadiniaz:2012xp, Bastianelli:2013pta, Ahmadiniaz:2015xoa, 
Ahmadiniaz:2015kfq, Edwards:2017bte, Ahmadiniaz:2017rrk}
for recent applications to gauge theories, refs.
\cite{Bastianelli:2002fv, Bastianelli:2002qw, Bastianelli:2005vk, Bastianelli:2005uy, Bastianelli:2004zp, Hollowood:2007ku,
 Bastianelli:2008cu, Bastianelli:2013tsa} for extensions to curved spaces, 
 and refs. \cite{Bastianelli:2007pv, Bastianelli:2008nm, Corradini:2010ia, Bastianelli:2012bn, Bonezzi:2017mwr,
 Bonora:2018uwx} for applications to higher spin theories. 
 Our main motivation for the present paper is to search for simpler methods for improving the efficiency of 
 worldline calculations in curved spaces. The case of type-A trace anomalies is both a useful check 
 as well as an interesting issue to investigate.
 
Thus, in Section~\ref{sec:2} we present a quick review of the scalar particle, which we then extended to 
the $N=1$ supersymmetric version of the model.  In Section~\ref{sec:3} we compute the perturbative expansion of the 
path integral for periodic (antiperiodic) boundary condition for worldline bosons (fermions),
as appropriate for addressing one-loop quantities in worldline applications, 
and in Section~\ref{sec:4} we present an application of the simplified path integral to 
identify the type-A trace anomalies of a Dirac fermion (correcting a minor misprint
for the spinor trace anomaly in $d=12$ found in the literature). 
We verify our results for the anomalies with the zeta function and holographic formulas 
mentioned above. Eventually, we present our conclusions and outlook in Section~\ref{sec:5}. To make the presentation self-contained
we list in Appendix~\ref{app-A} the relevant formulas for various geometrical objects of maximally symmetric spaces in
Riemann normal coordinates, and in Appendix~\ref{app-B} we report  a list of the Wick contractions used in the main text.

\section{Construction}
\label{sec:2}

A nonrelativistic particle of unit mass in a curved $d$-dimensional space has a lagrangian that takes the form of a nonlinear sigma model
\be 
 L(x,\dot x)=\frac12 g_{ij}(x)\dot x^i\dot x^j
 \label{nonlinear}
\ee 
 and corresponding hamiltonian
 \be
H(x,p)= \frac12 g^{ij}(x)p_i p_j 
 \ee
where  $g_{ij}(x)$ is the metric in an arbitrary coordinate system, $\dot x^i=\frac{d x^i}{dt}$, and $p_i$ the momenta conjugated to $x^i$.
Canonical quantization produces a quantum hamiltonian 
\be
\hat H_\xi (\hat x,\hat p) = \frac12 g^{-\frac14} (\hat x) \hat p_i g^{\frac12} (\hat x) g^{ij}(\hat x)
\hat p_j g^{-\frac14} (\hat x) +\frac{\xi}{2} R(\hat x)
\ee
where hats denote operators.
Ordering ambiguities have been partially fixed by requiring background coordinate invariance, leaving only 
a possible nonminimal coupling proportional to the scalar curvature $R$ and parametrized by the coupling constant $\xi$.
For simplicity we set the coupling $\xi=0$ (minimal coupling).
Other couplings, such as the conformal  coupling $\xi=\frac{d-2}{4(d-1)}$ or the value $\xi=\frac14$ that allows for a 
$N=1$ supersymmetrization (it appears in the square of the Dirac operator), can be reintroduced  by adding 
a scalar potential.

To obtain the simplified path integral, one starts by studying the evolution operator in euclidean time $\beta$ (the heat kernel)
\be
\hat K(\beta)=e^{-\beta \hat H}
\ee
where $ {\hat H}\equiv \hat H_0$ is the hamiltonian operator with minimal coupling, which satisfies the heat equation
\be
 -\partial_\beta  \hat K(\beta) = \hat H \hat K(\beta)  \;, \qquad  \hat K(0) =  \mathbb{1} \;.
\ee
Using  position eigenstates 
\be 
\hat x^i|x\ra =x^i|x\ra ~, \qquad  \la x|x'\ra = \frac{\delta^{(d)}(x-x')}{\sqrt{g(x)}} ~,
 \ee
 and corresponding resolution of the identity 
\be
 \mathbb{1}  = \int d^d x \sqrt{g(x)}\, |x\ra \la x|  ~,
 \ee
 one constructs scalar wave functions $\psi(x) = \la x| \psi\ra $ for any vector $|\psi\ra$ of the Hilbert space.  
 In particular, the matrix element of the evolution operator  
  \be 
 K(x, x';\beta) = \la x| e^{-\beta \hat H} |x'\ra
\ee
gives a biscalar under change of coordinates, and the heat equation takes the form
  \be
-\partial_\beta K(x, x';\beta) = -\frac12 \nabla^2_x \,  K(x, x';\beta)  \;,\qquad 
K(x, x';0) = \frac{\delta^{(d)}(x- x')}{\sqrt{g(x)}}~,
\ee
where $\nabla^2_x$  is the scalar laplacian $\nabla^2 =\frac{1}{\sqrt{g}} \partial_i \sqrt{g} g^{ij} \partial_j $
acting on the $x$ coordinates. Its solution has a well-defined path integral representation 
in terms of the nonlinear sigma model, see for example \cite{Bastianelli:2006rx}. 
However, one can manipulate the heat equation to obtain a simplified equation admitting a path integral 
representation in terms of a {\em linear} sigma model.
This is done as follows. One first transforms the transition amplitude  into a bidensity 
\be
\overline K(x,x', \beta) = g^{\frac14}(x) K(x,x', \beta) g^{\frac14}(x') ~,
\ee
for which the heat equation takes the form
 \be
 -\partial_\beta \overline K(x, x';\beta)  = -\frac12 g^{\frac14}(x) 
\nabla^2_{x}  \Big ( g^{-\frac14}(x) \overline  K(x, x';\beta)  \Big ) \;, \qquad
 \overline K(x, x';0) = \delta^{(d)}(x- x')  \;.
\ee
Then, one evaluates the differential operator appearing on the right hand side of this equation
to obtain the identity
\be
 -\frac12 g^{\frac14} \nabla^2  \, g^{-\frac14} = -\frac12 \partial_i g^{ij} \partial_j +V_{0}~,
 \label{diff-op}
\ee
with derivatives that act through, except in the effective scalar potential  $V_0$ given by
\be
V_{0} = -\frac12  g^{-\frac14} \partial_i \sqrt{g} g^{ij} \partial_j g^{-\frac14}\;, 
\ee
where all derivatives stop after acting on the last function. 
Thus the heat equation reads 
\be
- \partial_\beta \overline K(x, x' ;\beta) =
\Big (-\frac12 \partial_i  g^{ij}(x) \partial_j +V_{0}(x)  \Big )\overline K(x, x';\beta) \;.
\label{hk3}
\ee
At this stage one restricts to maximally symmetric spaces, uses Riemann normal coordinates centered at the initial point $x'$, 
and realizes that the metric $g^{ij}(x)$, appearing in  the term $\partial_i g^{ij}(x) \partial_j $, may be replaced
by the constant metric $\delta^{ij}$. This chain of steps  brings one to consider the equivalent heat equation
\be
-\partial_\beta \overline K(x, x';\beta) =
\Big (-\frac12 \delta^{ij}\partial_i \partial_j +V_{0}(x)  \Big )\overline K(x, x';\beta) ~,
\label{simplified-heat-eq}
\ee
where the simplified hamiltonian operator $H=-\frac12 \delta^{ij}\partial_i \partial_j +V_{0}(x)$  is
interpreted as that of a particle on a flat space (in cartesian coordinates), 
interacting with an effective potential $V_{0}$ of quantum origin (it is proportional to $\hbar^2$ in arbitrary units).
Evidently, this last equation admits a path integral representation in terms of a linear sigma model.

The replacement of  $g^{ij}(x)$ with $\delta^{ij}$ is valid since by the hypothesis of maximal symmetry,
and using Riemann normal coordinates with origin at $x'$ (which then has a vanishing value of its coordinates, namely $x'^i=0$), 
one deduces that  $K(x, 0;\beta)$ can only depend on the radial coordinate
  \begin{align}
  r = \sqrt{  \delta_{ij} x^ix^j} 
  \end{align}
  since there is no other possible tensor that may be used to contract the indices of the coordinates $x^i$ to form a 
  scalar\footnote{Note that in Riemann normal coordinates the metric takes the form given in \eqref{A4}, so that 
  $r^2\equiv \delta_{ij} x^i x^j =g_{ij}(x) x^i x^j$.}.
Then, using the explicit form of the metric (see appendix \ref{app-A}, eqs. \eqref{A4}--\eqref{A5}), one indeed  verifies that 
\begin{align}
\partial_i ( g^{ij}(x) \partial_j \overline K(x, 0;\beta)) =
\delta^{ij}\partial_i\partial_j \overline K(x, 0;\beta) 
\end{align}    
i.e. the validity of eq. \eqref{simplified-heat-eq} (again, we recall that in Riemann normal coordinates centered at $x'$,
the coordinates of the origin are $x'^i=0$)---see refs.~\cite{Bastianelli:2017wsy,Bastianelli:2017xhi} for more details.

Thus, one ends up with the linear sigma model 
\be 
 L(x,\dot x)=\frac12 \delta_{ij} \dot x^i\dot x^j + V_0(x)
 \label{linear}
\ee 
that can be used instead of \eqref{nonlinear} in a path integral to evaluate the transition amplitude 
$\overline K(x, 0;\beta)$  between an initial point  $x'^i =0$ (taken as the origin of the Riemann normal coordinates, 
that must be used in this set-up) and a final point $x^i$ in euclidean time $\beta$. 
As the space is maximally symmetric,  it is in particular  homogeneous,
and the origin can be chosen in any desired point of the manifold. This just to point out that the initial point of the heat kernel 
$K(x, 0;\beta)$ can be kept arbitrary.

The effective potential can be explicitly evaluated on maximally symmetric spaces, 
and for a sphere of radius $a$ and mass parameter $M=\frac{1}{a}$ one finds
\bea
V_0(x) 
\eqa  -\frac12  g^{-\frac14} \partial_i \sqrt{g} g^{ij} \partial_j g^{-\frac14}  
\ccr
\eqa   \frac{(d-1)}{8}  \Biggl[  \frac{(d-5)}{4}  \left (\frac{f'(r)}{1+f(r)}\right)^2 + \frac{1}{1+f(r)} \left ( \frac{(d-1)}{r} f'(r) + f''(r) \right) \Biggr] 
 \ccr
\eqa
\frac{d(1-d)}{12} M^2
+ \frac{( d-1) (d-3)}{48}  \frac{\Big(5 (Mr)^2-3 +\left((Mr)^2+3\right) \cos (2 M r)\Big)}{r^2 \sin ^2 (Mr)}   \ccr 
\qquad
\label{eff-pot}
\eea
where we have used the explicit metric in \eqref{A4} together with eq. \eqref{A5}. Note also that the potential is an even function of the 
radial coordinate $r=\sqrt{ \delta_{ij}x^i x^j}$.  The simplified path integral based on the linear sigma model \eqref{linear}
has been tested and used extensively in perturbative calculation in \cite{Bastianelli:2017wsy, Bastianelli:2017xhi},
verifying its superior efficiency with respect to the equivalent path integral based on the nonlinear sigma model, 
used for example in \cite{Bastianelli:2001tb}.

Let us now turn to the supersymmetric version of the particle mechanics, identified by the (euclidean) lagrangian
\be
L= \frac12 g_{ij}(x)\dot x^i \dot x^j 
+ \frac12  \psi^a (\dot \psi_a + \dot x^i \omega_{iab}(x) \psi^b)
\label{nlsigma}
\ee
where $\psi^a$ are real Grassmann variables with flat indices, and $\omega_{iab}$ is the spin connection built from the vielbein $e_i^a$.
The fermionic variables $\psi^a$ are the supersymmetric partners of the coordinates $x^i$. 
Upon quantization they lead to operators that satisfy the anticommutation relations  
\begin{align}
\big\{ \hat \psi^a, \hat \psi^b\big\} =\delta^{ab}~,
\label{eq:Galg}
\end{align} 
a Clifford algebra 
which can either be represented by the usual Dirac gamma matrices ($\hat \psi^a = \frac{1}{\sqrt 2}\gamma^a$, with
$\{\gamma^a, \gamma^b\}=2\delta^{ab}$), or treated by a fermionic path integral---we refer 
to~\cite{Bastianelli:2002qw} and references therein 
for further details on this supersymmetric model, and on its use in worldline calculations for Dirac fermions in background gravity.
Here we  just recall that the spinning particle model was originally introduced in \cite{Berezin:1976eg, Barducci:1976qu, Brink:1976sz}.

In the subsequent discussion we find it more useful to start our analysis using the gamma matrices. 
The conserved quantum supersymmetric charge of the model is proportional to the Dirac operator, and reads 
\begin{align}
&\hat Q =  -\frac{i}{\sqrt 2} \slashed{\nabla}(\omega) =
-\frac{i}{\sqrt 2} \gamma^ a e_a^i(x)\Big( \partial_i +\frac 14 \omega_{iab}(x) \gamma^a \gamma^b\Big) 
\end{align} 
while the related quantum hamiltonian becomes
\begin{align}
&\hat H = \hat Q ^2 = -\frac 12 \slashed{\nabla}^2 =-\frac12 g^{ij}(x)\nabla_i(\omega,\Gamma) \nabla_j(\omega) + \frac18 R~,
\label{eq:H1/2}
\end{align}
where we have indicated the connections present in the various covariant derivatives.
Of course, all these operators  act on a spinorial wave function (a Dirac spinor).

The heat kernel associated to this hamiltonian 
\be
{\mathbb K} = e^{-\beta \hat H}
\ee
has quantum mechanical matrix elements 
\be
{\mathbb K}_{\alpha\alpha'}(x,0;\beta) =\langle x,\alpha| e^{-\beta \hat H}|0,\alpha'\rangle \nonumber\\
\ee
where $\alpha$, $\alpha'$  are spinorial indices.
In the following we will not show the spinorial indices explicitly, and just remember that ${\mathbb K}$ is matrix-valued.
Now, using  the fact that the space under consideration is maximally symmetric, one deduces that 
the heat kernel ${\mathbb K}(x,0;\beta)$ can only be a function of  
 $x^2$, $\gamma_a \gamma^a\sim {\mathbb 1}$ and $\delta_{ia} x^i \gamma^a$. 
In addition,  as the gamma matrices appear only in even combinations (they are contained quadratically in the spin connections 
inside the hamiltonian \eqref{eq:H1/2}), one finds that  the dependence on $\delta_{ia} x^i \gamma^a$ arises only 
through its square
\be
(\delta_{ia} x^i \gamma^a)^2 =  {\mathbb 1}\,  x^2
\ee
which is again proportional to the identity matrix.
Thus, the full heat kernel is proportional to the identity, and must be a function of $r=\sqrt{ \delta_{ij}x^i x^j}$ only,
\begin{align}
{\mathbb K}(x,0;\beta) = {\mathbb 1}\, U(r ;\beta)~.
\end{align}    
 
Equipped with this result let us analyze the heat equation satisfied by the bidensity
\be
\overline {\mathbb K}(x,0;\beta) = g^{1/4}(x) {\mathbb K}(x,0;\beta)  g^{1/4}(0)~,
\ee
(the value $g(0)=1$ is actually irrelevant) which is 
\begin{align}
-\partial_\beta \overline {\mathbb K}(x,0;\beta)  = g^{1/4}(x) \hat H g^{-1/4}(x)\, \overline {\mathbb K}(x,0;\beta) ~.
\label{eq:K1/2}
\end{align}
By expanding out the expression of the hamiltonian given in equation~\eqref{eq:H1/2}, we write 
\begin{align}
g^{1/4} \hat H g^{-1/4} =&-\frac12 g^{1/4} \nabla^2 g^{-1/4} \nonumber\\
&-\frac18 (\partial_i \omega^i{}_{ab}) \gamma^{ab}-\frac14 \omega^i{}_{ab}\gamma^{ab} \partial_i 
\nonumber\\
&-\frac1{32}\omega_{iab}\, \omega^i{}_{cd}\, \gamma^{ab}\gamma^{cd}+\frac18 R~.
\label{eq:Hexp}
\end{align} 
Using the explicit expression of the spin connection \eqref{eq:O-RN}, which satisfies the Fock-Schwinger gauge~\eqref{eq:FS}, it is easy to check that the terms in the second line do not contribute when applied to $\bar {\mathbb K}$
(they give rise to terms proportional to $x^i \omega_{iab}\sim 0$, because of the Fock-Schwinger gauge),
whereas the terms in the third line give expressions proportional to ${\mathbb 1}$.  In particular we find
\begin{align}
-\frac1{32} \omega_{iab}\, \omega^i{}_{cd}\, \gamma^{ab}\gamma^{cd} = \frac{d-1}{8} M^2 \left( \frac{1-\cos(Mr)}{\sin(Mr)}\right)^2 {\mathbb 1}~.
\end{align} 
Thus, recalling  eq. \eqref{diff-op} and the possibility of replacing  $g^{ij}(x)$ with $\delta^{ij}$ in the first term of  \eqref{diff-op},
 we find  that a simplified heat equation holds
\begin{align}
-\partial_\beta \overline {\mathbb K}(x,0;\beta) 
=\Big( -\frac12 \delta^{ij}\partial_i \partial_j +V_{\frac12}(x) \Big) \overline {\mathbb K}(x,0;\beta)
\label{eq:heat1:2}
\end{align}
with
\begin{align}
V_{\frac12}(x) =V_{0}(x) + \frac{d(d-1) M^2}{8} +\frac{d-1}{8} M^2 \left( \frac{1-\cos(Mr)}{\sin(Mr)}\right)^2
\label{eq:V1:2}
\end{align}
where the second addendum is just $\frac18 R$, and $V_{0}(x)$ as given in \eqref{eff-pot}.

Expressions~\eqref{eq:heat1:2} and~\eqref{eq:V1:2} are the crucial results. They allow to find a simplified path integral.
 The message they carry is that the heat kernel of a spinorial operator on maximally symmetric spaces, when written in Riemann normal coordinates, satisfies a flat heat equation with the information on the curvature fully encoded in an  effective potential, just as it happens for the heat kernel of a scalar particle. As such it is straightforward to represent it as the path integral 
 of a linear sigma model
\begin{align}
\overline {\mathbb K} (x,0;\beta) = {\mathbb 1}\int_{x(0)=0}^{x(\beta)=x}Dx~e^{-S[x]}\,,\quad S[x] =\int_0^\beta dt\Big( \frac12 \delta_{ij} \dot x^i(t) \dot x^j(t) + V_{\frac12}(x(t))\Big) 
\label{eq:HKf}
\end{align} 
with the effective potential $V_{\frac12}$ given explicitly as a function of $r=\sqrt{ \delta_{ij}x^i x^j}$  by
\bea
V_{\frac12}(x)\eqa \frac{d(d-1)}{24} M^2
+ \frac{( d-1) (d-3)}{48}  \frac{\Big(5 (Mr)^2-3 +\left((Mr)^2+3\right) \cos (2 M r)\Big)}{r^2 \sin ^2 (Mr)}  \ccr
&+& \frac{( d-1)}{8} M^2 \left (\frac{1 -\cos(Mr)}{\sin(Mr)} \right )^2 ~.
   \eea
   
Of course, one could reintroduce free worldline fermions $\psi^a$ to represent the identity with a Grassmann path integral,
so to have the full linear sigma model lagrangian
    \be
L= \frac12 \delta_{ij} \dot x^i \dot x^j + \frac12 \psi_a\dot \psi^a +  V_{\frac12}(x) 
   \ee
which may be compared with the original nonlinear  sigma model we started with in eq.~\eqref{nlsigma}.
One could then use antiperiodic boundary conditions on the $\psi$'s to produce the trace on the spinor indices,
periodic boundary conditions to produce the trace with an insertion of $\gamma^5$, or more generally 
leave open boundary conditions. However,  at this stage this is just an amusing observation,
as the heat kernel remains trivial on the spinor indices, in particular traces are trivially computed.

In the following section we test the previous simplified path integral by computing 
its perturbative expansion. We then use it to obtain the type-A trace anomalies of a Dirac field coupled to gravity in dimensions $d\leq 16$.  

\section{Perturbative expansion} 
\label{sec:3}
The short-time perturbative expansion of the kernel~\eqref{eq:HKf} can be formally written as a power series in $\beta$
\begin{align}
\overline{\mathbb K} (x,0;\beta) = g^{1/4}(x) \frac{ e^{-\frac{x^2}{2 \beta}} }{(2\pi\beta)^\frac d2}  
\sum_{n=0}^\infty a_n(x,0) \beta^n ~,
\label{eq:HKexp}
\end{align}
where $a_n$ are the so-called Seeley-DeWitt coefficients. 
In general they are matrix-valued, but as we have discussed they are proportional to the identity
matrix on maximally symmetric spaces. 
In order to compute perturbatively the  expansion with our simplified path integral, we find it  convenient to use a  rescaled  time $\tau=t/ \beta$, so that 
\begin{align}
S[x] =\int_0^1 d\tau\Big( \frac1{2\beta} \delta_{ij} \dot x^i \dot x^j + \beta V_{\frac12}(x(\tau))\Big)~, 
\end{align}
where the dot now indicates derivative with respect to $\tau$. Then we Taylor expand the potential about the origin
of the Riemann coordinates 
\begin{align}
V_{\frac12}(x) =M^2\sum_{l=0}^\infty k_{2l}\, (Mr)^{2l}\,, \quad \Rightarrow\quad S[x]=\int_0^1 d\tau \frac1{2\beta} \delta_{ij} \dot x^i \dot x^j +\sum_{l=0}^\infty S_{2l}[x]
\end{align} 
and retain only the relevant ``coupling constants'' $k_{2l}$ needed to carry out the expansion at the desired order. 
Explicitly,
\be
S_{2l}[x] = \beta M^{2+2l} k_{2l} \int_0^1 d\tau\,  (\delta_{ij}x^i x^j)^l \;. 
\ee
The perturbative expansion is obtained by considering that the propagator associated to  the free kinetic term is of order $\beta$, and reads
\begin{align}
\langle x^i(\tau) x^{j}(\tau') \rangle = -\beta \delta^{ij} \Delta(\tau,\tau')\,,\quad  \Delta(\tau,\tau')=\frac12 |\tau -\tau'| -\frac12 (\tau+\tau') +\tau \tau'~,
\end{align} 
while each vertex adds a power of $\beta$. Therefore, in order to carry out an expansion say to order $\beta^m$, one needs to retain couplings up to $k_{2(m-1)}$. 

Specifically, we compute the expansion up to order $\beta^8$, which requires the following coupling constants 
extracted from $V_{\frac12}$~\footnote{The first addenda in the expressions 
below are the contributions from $V_0$, the others are the contribution from $\frac18 R$ and $\omega \omega$ term.}
{\allowdisplaybreaks
\bea
k_0 \eqa d(d-1)\left(-\frac1{12} + \frac18\right) = \frac{d(d-1)}{24} 
\label{eq:ccs}\\[2mm]
k_2 \eqa (d-1) \left ( (d-3) \frac{1}{120} + \frac{1}{32}\right)  = \frac{(d-1) (4 d+3)}{480}
\ccr[2mm]
k_4 \eqa (d-1)  \left ((d-3) \frac{1}{756}   +   \frac{1}{192} \right)  =
\frac{(d-1) (16 d+15)}{12096}
\ccr[2mm]
k_6 \eqa (d-1) \left ( (d-3) \frac{1}{5400} + \frac{17}{23040}
 \right) 
= \frac{(d-1) (64 d+63)}{345600}
\ccr[2mm]
k_8 \eqa (d-1) \left ( (d-3) \frac{1}{41580} + \frac{31}{322560}
\right)
=\frac{(d-1) (256 d+255)}{10644480}
\ccr[2mm]
k_{10} \eqa (d-1) \left ( (d-3) \frac{691}{232186500} + \frac{691}{58060800}
\right)
=\frac{691 (d-1) (1024 d+1023)}{237758976000} 
\ccr[2mm]
k_{12} \eqa (d-1)  \left ( (d-3)  \frac{1}{2806650} + \frac{5461}{3832012800}
\right)
= \frac{(d-1) (4096 d+4095)}{11496038400}
\ccr[2mm]
k_{14} \eqa (d-1) \left (  (d-3) \frac{3617}{86837751000} + \frac{929569}{5579410636800}
\right)
= \frac{3617 (d-1) (16384 d + 16383)}{1422749712384000}~.\nonumber
\eea
}
For simplicity, we consider  the diagonal part of the heat kernel only by setting $x=0$, which is relevant to obtain the trace anomalies
or to compute the one-loop effective action of a Dirac spinor. This involves the following correlators
{\allowdisplaybreaks
\bea
&&\overline {\mathbb K}(0, 0;\beta)  = \frac{{\mathbb 1}}{(2\pi \beta)^{\frac{d}{2}}}  e^{-S_0} 
\exp \biggl [ - \underbrace{\langle S_2 \rangle}_{O(\beta^{2})}
-\underbrace{\langle S_4 \rangle}_{O(\beta^{3})}
\ \underbrace{ - \langle S_6 \rangle +\frac12  \langle S_2^2 \rangle_c}_ {O(\beta^{4})}
\ \underbrace{- \langle S_8 \rangle + \phantom{\frac12}\hskip-.35cm\langle S_4 S_2 \rangle_c }_ {O(\beta^{5})}
\ccr
&& 
\underbrace{ - \langle S_{10} \rangle + \langle S_6 S_2 \rangle_c  
+\frac12   \langle S_4^2 \rangle_c  -\frac{1}{3!}  \langle S_2^3 \rangle_c}_{O(\beta^{6})}
\ \underbrace{- \la S_{12} \ra   +
\la S_8 S_2\ra_c  +
\la S_6 S_4\ra_c  -\frac12
\la S_4 S_2^2 \ra_c}_{O(\beta^{7})}
\ccr
&&
 \underbrace{-\la S_{14}\ra  
+\la S_{10} S_2\ra_c 
+\la S_8 S_4\ra_c 
+\frac12 \la S_6^2 \ra_c 
-\frac12 \la S_6 S^2_2\ra_c 
-\frac12 \la S_4^2 S_2\ra_c 
+\frac{1}{4!} \la S_2^4\ra_c}_{O(\beta^{8})}
\ccr
&&
+ O(\beta^{9}) \biggr ] 
\label{eq:dkernel}
\eea
}
where the subscript ``$c$" stands for ``connected" correlation functions.

Previously, in refs.~\cite{Bastianelli:2017wsy, Bastianelli:2017xhi} the same set of correlators for the scalar heat kernel
was computed. 
The expression for the kernel~\eqref{eq:dkernel} differs from that obtained in the scalar case only in the coupling constants, 
now given by~\eqref{eq:ccs}. Hence, the final result for the fermion heat kernel at coinciding points can be obtained by plugging
 the new coupling constants into the expression of the scalar heat kernel, 
reported in Appendix~\ref{app-B} for completeness. Thus we get  
{\allowdisplaybreaks
\begin{align}
&\overline {\mathbb K}(0, 0;\beta)  =
\frac{{\mathbb 1}}{(2\pi \beta)^{\frac{d}{2}}} 
 \exp \biggl [- d(d-1)\frac{\beta M^2 } {24}  
 + d(d-1)\biggl( -\frac{(\beta M^2)^2}{6!} \frac{ 4 d+3}{4} 
 \ccr&
 - \frac{(\beta M^2)^3}{9!}  (d+2) (16 d+15) \ccr&
 - \frac{(\beta M^2)^4}{10!}
   \frac{16 d^3 + 257 d^2 + 555 d +315}{8}
  \ccr & 
+ \frac{(\beta M^2)^5}{11!}  \frac{ (d+2) \left(64 d^3 -333 d^2 -1341 d -945 \right)}{24} 
\ccr
&
+
\frac{(\beta M^2)^6}{13!} \frac{207744 d^5 +943595 d^4 -2652226 d^3 -18403426 d^2 -29381262d
-14365890}{5040}
\ccr
   &
+ \frac{(\beta M^2)^7}{14!} 
\frac{(d+2) \left( 16896 d^5 +243703 d^4 +213650 d^3 -2640054 d^2 -6680970 d -4054050
\right)}{720}
   \ccr
  &
- \frac{(\beta M^2)^8}{17!} \tfrac{3175680 d^7  -132047423 d^6 -1198310651 d^5 -2099217371 d^4 +8069209407d^3
+36235883583 d^2 +49125794355 d +21995248275}{1440}
   \ccr
  &
 + O(\beta^{9}) 
 \biggl) \biggr]
 \label{eq:K00}
\end{align}
}
which could equivalently be written in terms of the constant scalar curvature  $R$.
In this expression the exponential must be expanded, keeping terms up to order  $O(\beta^{8})$ included. This allows to
read off the diagonal coefficients $a_n(0,0)$, with integer $n$ up to $n=8$. We use them in the next section 
to identify the type-A trace anomaly of a Dirac fermion in various dimensions.

\section{The type-A trace anomalies}
\label{sec:4} 

The path integral calculation of the transition amplitude on a maximally symmetric space can be employed and tested
to evaluate the type-A trace anomaly of a massless Dirac fermion coupled to gravity, in space-time dimensions~$d\leq~16$. 
These anomalies are the ones proportional to the topological Euler density
of the curved space \cite{Deser:1993yx}, see also \cite{Boulanger:2007ab, Boulanger:2007st}
for their cohomological characterization.
 
In general, the trace anomaly of a Dirac fermion
can be related to the transition amplitude of a $N=1$ spinning particle in a curved space by 
\begin{align}
\big\langle T^m{}_m(x)\big\rangle_{QFT} = - \lim_{\beta\to 0} {\rm tr}\, {\mathbb K}(x,x;\beta)~,
\label{trace-an}
\end{align}
where on the left hand side $T^m{}_m(x)$ is the trace of the  stress tensor of the Dirac spinor
in a curved background, obtained from the appropriate Dirac action $S_{D}$ by
$T_{m a}(x) = \frac{1}{e}\frac{\delta S_{D}}{\delta e^{m a} (x)} $
where $e_m^a (x)$ is the vielbein of the curved spacetime.
The expectation value is performed in the corresponding quantum field theory.
The right hand side can be viewed as the anomalous contribution arising from the
QFT path integral measure, regulated {\it \`a la} Fujikawa~\cite{Fujikawa:1980vr},
with the minus sign being the usual one due to the fermionic measure, and the trace
being the trace on spinor indices.    
The regulator corresponds to the square of the Dirac operator, and is identified with the quantum hamiltonian $\hat H$ 
of the  $N=1$ spinning particle in a curved space
\be
\hat H=  -\frac12 ({\slash \hskip -2.3mm\nabla})^2~,
\label{eq:nabla2}
\ee 
which appears in the heat kernel at coinciding points $\mathbb{K}(x,x;\beta)$. The latter can be evaluated with 
a path integral \cite{Bastianelli:1991be, Bastianelli:1992ct}. It  is understood that the $\beta \to 0$ limit  
in \eqref{trace-an} picks up just the $\beta$-independent term, as divergent terms are removed 
by the QFT renormalization.  This procedure selects the appropriate heat kernel coefficient $a_n(x,x)$ sitting 
in the expansion of  ${\mathbb K}(x,x;\beta)$. It may be interpreted as the contribution to the anomaly of the 
regularized particle making its virtual loop, see for example 
\cite{Diaz:1989nx}, where a Pauli-Villars regularization  gives rise to the Fujikawa regulator used above.

Expanding ${\mathbb K}(x,x;\beta)$ at the required order one can read off the trace anomalies in even $d$ dimensions
(odd dimensions support no anomaly if the space is boundaryless)
\begin{align}
\big\langle T^m{}_m(x)\big\rangle_{QFT} = -\frac{ {\rm tr}\, a_\frac{d}{2}(x,x)}{(2 \pi)^\frac{d}{2}}
\label{trace-an1}
\end{align}
that is, for even $d=2n$ dimensions, the relevant coefficient is precisely  $a_n(x,x)$.
Of course, one may use Riemann normal coordinates centered at $x$, so that $\sqrt{g(x)}=1$ and 
$\overline {\mathbb K}(x,x;\beta) = {\mathbb K}(x,x;\beta)$. This formula holds on a generic space. 
In the present maximally symmetric case, due to translational invariance, the choice of which point is the origin 
of the Riemann coordinates becomes irrelevant.  Hence, $\overline {\mathbb K}(x,x;\beta) =\overline {\mathbb K}(0,0;\beta) $, 
and the result obtained in the previous section is directly applicable. The trace in \eqref{trace-an1}
reduces to the trace of the identity matrix, and counts the dimension of the spinor space, 
$2^\frac{d}{2}$ for even $d$ dimensions.

In Table~\ref{table}, we list the anomalies we obtain from the expansion~\eqref{eq:K00},
expressing the results both in terms of the scalar curvature $R$ and in terms of the sphere radius $a=\frac{1}{M}$.

{ 
\renewcommand{\arraystretch}{2}
\begin{table}[!ht]
\begin{center}
\begin{tabular}{ | l  | p{7cm} |    p{4cm} | }
    \hline 
  $d$  &   $\la T^m{}_m\ra$  &  $\la T^m{}_m\ra$  \\   \hline
2 & \large $\frac{R}{24\, \pi} $ 
& \large 
$\frac{1}{12\, \pi  a^2 }  $ 
\\ \hline
4 & \large $-\frac{11\, R^2}{34\, 560\, \pi^2} $ 
& \large 
$-\frac{11}{240\, \pi ^2 a^4}  $ 
\\ \hline
6 & \large $\frac{191\, R^3}{108\, 864\, 000\, \pi^3} $  
& 
\large $\frac{191}{4032\, \pi ^3 a^6} $
\\ \hline
8 & \large  $-\frac{2\,497\, R^4}{339\, 880\, 181\, 760\, \pi^4} $   
&\large  $-\frac{2\,497}{34\,560\, \pi ^4 a^8}$  
\\ \hline
10 & \large  $\frac{14\,797\, R^5}{598\, 615\, 142\, 400\, 000\, \pi^5} $   
& \large $\frac{14\,797}{101\,376\, \pi ^5 a^{10} }$
 \\ 
\hline
12 & \large $-\frac{92\,427\,157\, R^6}{1\, 330\, 910\, 037\, 208\, 675\, 123\, 200\, \pi^6}  $
& \large $-\frac{92\,427\,157}{251\,596\,800 \pi ^6 a^{12} } $
\\
\hline
14 & \large $\frac{36\,740\,617 \, R^7}{219\,454\,597\,066\,612\,329\,676\,800\,
\pi^7}  $
& \large $ \frac{36\,740\,617}{33\,177\,600\, \pi ^7 a^{14}} $
\\
\hline
16 & \large $-\frac{61\,430\,943\,169 \, R^8}{173\, 836\, 853\, 795\, 629\, 301\, 760\, 000\, 000\, 000\,
\pi^8} $
& \large  $-\frac{61\,430\,943\,169}{15\,792\,537\,600\, \pi ^8 a^{16}}$
\\
\hline
    \end{tabular}
    \caption{Type-A trace anomaly of a Dirac spinor in terms of the curvature scalar $R$, and in terms of the radius $a$, in various dimensions.
~\label{table}}
    \end{center}
    \end{table}
}
These anomalies were listed up to $d=12$ in terms of the radius $a$ also in ref. \cite{Copeland:1985ua} 
(however the value of the anomaly in $d=12$ reported there is incorrect, their denominator differs from ours,
presumably a misprint).

 The type-A trace anomaly can also be obtained using the Riemann zeta-function associated to the differential 
 operator~\eqref{eq:nabla2} 
\begin{align}
\big\langle T^m{}_m(x)\big\rangle_{QFT} =
 -\frac{\Gamma\left(\frac{d+1}{2}\right)}{2\pi^{\frac{d+1}{2}} a^d}\, \zeta_{{\slash{\hskip-2mm\nabla}}^2}(0)~,
 \label{t-an}
\end{align}
as discussed in~\cite{Copeland:1985ua, Cappelli:2000fe}. More recently, an efficient way of computing such trace anomalies within the AdS/CFT correspondence was proposed in~\cite{Aros:2011iz}. 
In that reference, a simple formula for certain coefficients $c^{(d)}_{_{{\slash{\hskip-1.75mm\nabla}}^2}} $
linked to the Riemann zeta function was found
\begin{align}
c^{(d)}_{_{{\slash{\hskip-1.75mm\nabla}}^2}} = \frac{4 (-1)^{\frac d2}}{(8 \pi)^{\frac d2} \left(\frac d2\right)! \left(\frac{d}2 -1\right)!} 
\int_0^{\frac12} d\nu \frac{\left(\frac12 +\nu\right)_{\frac d2} \left(\frac12 -\nu\right)_{\frac d2}}{\left(\frac12\right)_{\frac d2}}~,
\end{align}  
where $(x)_n = \frac{\Gamma(x+n)}{\Gamma(x)} = x (x+1) \dots (x+n-1)$ is the Pochhammer symbol (the raising factorial).
We have checked that these coefficients are linked explicitly to the Riemann zeta function by
\begin{align}
\zeta_{{\slash{\hskip-2mm\nabla}}^2}(0)=  (4 \pi)^{\frac d2} \left( \frac d2 -1\right)! ~c^{(d)}_{_{{\slash{\hskip-1.75mm\nabla}}^2}}~.
\end{align}
which, in turn, allow to identify the anomaly in  \eqref{t-an}.
One can check that  both methods reproduce the same type-A trace anomalies, which indeed coincide with the ones computed 
by the simplified path integral in $d=2,...,16$ and listed in Table~\ref{table}. 

Finally, it is convenient to summarize the type-A trace anomalies by presenting them in the form
\begin{align}
\big\langle T^m{}_m(x)\big\rangle = (-1)^{n+1} a_{2n} \frac{E_{2n}}{(2\pi)^n} 
\end{align}
where $E_{2n}$ is the Euler density of the $d=2n$ dimensional space defined by
\be
E_{2n} = \frac{(2n)!}{2^n} R_{m_{_1} m_{_2}}{}^{[m_{_1}m_{_2}} \dots R_{m_{_{2n-1}} m_{_{2n}}}{}^{m_{_{2n-1}}m_{_{2n}}]}
\ee
(the square bracket denotes weighted antisymmetrization) with $a_{2n}$ the constant anomaly coefficient. 
The stress tensors are normalized as usual,  
$T_{mn} = \frac{2}{\sqrt{g}} \frac{\delta S}{\delta g^{mn}}$ for the scalar and 
$T_{ma} = \frac{1}{e} \frac{\delta S}{\delta e^{ma}}$ for the spinor, where $g_{mn}$ and $e_{ma}$ 
are the metric and vielbein, respectively. The sign $(-1)^{n+1}$ is conventionally inserted to make 
the coefficients $a_{2n}$ positive, as we will check shortly.
On spheres the Euler density  evaluates to
\be
E_{2n} = \frac{(2n)!}{(2n(2n-1))^n} R^n
\ee
and from the previous discussion we identify the following coefficients for a Dirac fermion 
\be
a_{2n}^{fermion} =   \frac{2}{n!(2n)!}   
\int_0^{\frac12} dx\, \prod_{i=0}^{n-1} \left( \left(i+\frac12\right)^2 - x^2 
\right) \;.
\ee
Similarly, we also identify the coefficients  for a real conformal scalar, 
using formulas from \cite{Diaz:2008hy}, 
\be
a_{2n}^{scalar} = -  \frac{(2n-1)!!}{((2n)!)^2} \int_0^1 dx \prod_{i=0}^{n-1} (i^2 - x^2) \;.
\ee
By inspection, one may notice that these coefficients are positive for every $n$, as the integrands are products of positive functions in the given 
range of integration (for the scalar, the explicit minus sign makes positive the contribution of the $i=0$ term of the product).
This positivity is not evident in the worldline method, and appears only at the end of our calculations.
In general, these coefficients are expected to be positive, as they appear in conjectured higher dimensional extensions of the $a$-theorem
and interpreted as a measure of the effective degrees of freedom at the fixed point. These conjectured  $a$-theorems
extend suitably the $c$-theorem of two dimensions \cite{Zamolodchikov:1986gt}
and the $a$-theorem of four dimension \cite{Komargodski:2011vj}, where indeed the coefficients have been proven 
to be positive for arbitrary unitary conformal field theories. For $d=2n=2,4,..., 16$ we report the values of these coefficients, as well as their ratio \cite{Cappelli:2000fe}, in Table \ref{table2}.
{ 
\renewcommand{\arraystretch}{2}
\begin{table}[!ht]
\begin{center}
\begin{tabular}{ | l  | p{3.8cm} |    p{3.8cm} |  p{2cm} | }
    \hline 
  $d=2n$  &   $(2n+1)!$ $a(scalar)$ &  $ (2n+1)!$ $a(fermion)$  
& $\frac{a(fermion)}{a(scalar)}$   \\   
  \hline
2 &  $\frac{1}{2}$ &  $\frac{1}{2} $ & $ 1$ 
\\ \hline
4 & $ \frac{1}{12} $ & $\frac{11}{12} $  & $11$ 
\\ \hline
6 & $  \frac{5}{72}$ & $  \frac{191}{72}$ & $\frac{191}{5}$
\\ \hline
8 &  $\frac{23}{240} $ &  $\frac{2497}{240}$  & $\frac{2497}{23}$
\\ \hline
10 &  $\frac{263}{1440} $ &  $\frac{14797}{288}$ & $\frac{73985}{263}$
 \\ 
\hline
12 &  $ \frac{133787}{302400} $ &   $\frac{92427157}{302400} $ 
& $\frac{92427157}{133787} $
\\
\hline
14 & $ \frac{157009}{120960} $
& $ \frac{36740617}{17280}$ & $\frac{257184319}{157009}$
\\
\hline
16 & $\frac{16215071}{3628800}$
& $ \frac{61430943169}{3628800} $ & $\frac{61430943169}{16215071}$
\\
\hline
    \end{tabular}
    \caption{The $a$ coefficients of the type-A trace anomaly of a real conformal scalar and  Dirac fermion.
    We have multiplied them by $(2n+1)!$ to make the numbers more readable.
    ~\label{table2}}
    \end{center}
    \end{table}
}

\section{Conclusions}
\label{sec:5}
We have considered the worldline path integral for the $N=1$ supersymmetric quantum mechanics in curved space, which is characterized 
by a supersymmetric non-linear sigma model action. We have shown that, when the space has maximal symmetry, 
the nonlinear sigma model can be traded with a purely bosonic linear sigma model and the curvature effects are taken care of by a suitable
effective scalar potential, which extends the one studied in~\cite{Bastianelli:2017wsy,Bastianelli:2017xhi}. 
We have tested our model by computing  the type-A trace anomalies of a Dirac fermion in space-time dimension $d\leq 16$, showing that they 
match those obtained with other techniques. However, further checks can be performed. 
Firstly, since the full heat kernel for maximally symmetric spaces  in Riemann normal coordinates is found to be proportional to the spinorial identity, 
the gravitational contribution to the chiral anomaly results proportional to  ${\rm tr} \,\gamma^5$, and thus correctly vanishes.  
On the other hand, we can also verify that, for $d=3,5$, the expansions of the diagonal heat kernels have vanishing coefficients $a_n$ for $n\geq 2,3$ 
respectively, as predicted by Camporesi in his exact calculations of spinorial heat kernels in maximally symmetric spaces~\cite{Camporesi:1992tm}.   

The present construction can presumably be extended also to the $N=2$ supersymmetric quantum mechanics 
used in the description of differential $p$-forms and particles of spin 1
\cite{Howe:1989vn, Bastianelli:2005vk, Bastianelli:2005uy},
as well as to the  supersymmetric quantum mechanics at arbitrary $N$, which provide the degrees of freedom of 
the first quantized approach to higher spinning particles \cite{Gershun:1979fb, Howe:1988ft}. The latter
enjoy conformal symmetry  \cite{Siegel:1988ru, Siegel:1988gd},
and  can be coupled to maximally symmetric spaces \cite{Kuzenko:1995mg}
(more generally, to conformally flat spaces \cite{Bastianelli:2008nm}).
The path integral for the spinning particle with $N$ supersymmetries  
on curved spaces needs regularization schemes with suitable regularization-dependent counterterms
\cite{Bastianelli:2011cc}. A linear sigma model approach may simplify drastically the situation, at least 
on maximally symmetric spaces.

Finally, a direction  worth looking at is the inclusion of boundaries in the maximally symmetric 
spaces, extending to curved spaces the worldline treatments 
of refs. \cite{Bastianelli:2006hq, Bastianelli:2008vh}, and study in particular 
the possible trace anomalies supported by the boundaries, as discussed for example in 
\cite{Solodukhin:2015eca, Fursaev:2016inw, Rodriguez-Gomez:2017kxf, Rodriguez-Gomez:2017aca}.

\vfill\eject

\appendix   
\section{Riemann normal coordinates in maximally symmetric spaces}
\label{app-A}

Maximally symmetric spaces are spaces with a maximal number of isometries.
Their curvature tensors can be expressed in terms of the metric as  
\bea
R_{ijmn} \eqa M^2 (g_{im}g_{jn}-g_{in}g_{jm} ) \\[2mm]
R_{ij} \eqa R_{mi}{}^m{}_j = M^2 (d-1) g_{ij}  \\[2mm]
R \eqa R_{i}{}^i =
M^2 (d-1) d ~,
\eea
where  $M^2=1/a^2$ is a constant, which is positive for a  sphere of radius $a$, 
vanishes for a flat space, and  is negative for a real hyperbolic space. 
This exhausts the list of simply connected, maximally symmetric spaces. 
For simplicity we consider spheres, as real hyperbolic spaces can be obtained by a simple analytic continuation.

In the main text we use Riemann normal coordinates (for details see \cite{Eisenhart:1965, Petrov:1969, Muller:1997zk}), 
and \cite{AlvarezGaume:1981hn, Howe:1986vm, Bastianelli:2000dw} for applications to nonlinear sigma models).
On spheres the sectional curvature is positive, and we can take $M=\frac{1}{a} >0$. One may then   
evaluate recursively all terms in the expansion of the metric  and sum them up \cite{Bastianelli:2001tb}, to obtain
\be
 g_{ij}(x) =  \delta_{ij} +f(r) P_{ij} 
 = \delta_{ij} + f(r) P_{ij}  \label{A4}
\ee
where $x^i$ denote the Riemann normal coordinates centered around a point (the origin), 
$P_{ij}$  indicates a projector given by
\be
P_{ij} = \delta_{ij} - \hat x_i \hat x_j ~,  \qquad  \hat x^i = \frac{x^i}{r} ~, \qquad  
r = \sqrt{\vec x^{\, 2}} ~,
\label{A4-bis}
\ee
and
\be
f(r) = \frac{1-2 (Mr)^2 - \cos(2Mr)}{ 2 (Mr)^2}~. \label{A5} 
\ee
Note that the function $f(r)$ does not have poles and it is even in $r$, so that it depends only on 
$r^2=\vec x^{\, 2} =\delta_{ij}x^i x^j$.  Note also that, because of the projector $ P_{ij}$
one has the equality  $r^2=g_{ij}(x)  x^i x^j$.
The inverse metric $ g^{ij}(x)$ and metric determinant   $g(x)$  are given by
\bea
 g^{ij}(x) \eqa \delta^{ij}  -\frac{f(r)}{1+f(r)}
 P^{ij}  \label{A8} \\[2mm]
  g(x)\eqa (1+f(r))^{d-1} ~.
\label{eq:g-RNC}
\eea

It is easy to check that the metric in~\eqref{A4} can be generated by the following choice of vielbein
\begin{align}
e^a_i(x) =\delta^a_i+l(r)P^a_i(x)
\label{eq:viel}
\end{align} 
where $x^a = \delta^a_i x^i$ and~\footnote{A priori, there are two independent solutions $l_\pm(r) =-1\pm \sqrt{1+f(r)}$ of the quadratic equation that follows from $g_{ij} = \eta_{ab}e^a_i e^b_j$. However, only with the upper solution does the vielbein reduce to the flat vielbein  when $M^2\to 0$.}
\begin{align}
l(r) =-1+\sqrt{1+f(r)} = -1+\frac{\sin(Mr)}{Mr}~.
\label{eq:l-MS}
\end{align}
The inverse vielbein reads instead 
\begin{align}
e^{ai}(x) =\delta^{ai}+\left( -1+\frac{1}{\sqrt{1+f(r)}}\right)P^{ai}(x)~.
\end{align}
Thus, by using the relation
\begin{align}
\omega_i{}^{ab}(x)= \frac12 e^{aj}\big( \partial_i e^b_j -\partial_j e^b_i\big)-\frac12 e^{bj}\big( \partial_i e^a_j -\partial_j e^a_i\big)
-\frac12 e^c_ie^{aj}e^{bk}\big( \partial_j e_{ck} -\partial_k e_{cj}\big)
\end{align}
one can promptly compute the associated spin connection, which is found to be
\begin{align}
\omega_i{}^{ab}(x)=\Omega(r) \frac12 x^j\big(\delta_j^a\delta_i^b- \delta_j^b\delta_i^a\big) 
\label{eq:ohm}
\end{align}
with 
\begin{align}
\Omega(r) = -\frac{2}{r} \left(l'(r)+\frac{l(r)}{r}\right) = 2M^2 \frac{1-\cos(Mr)}{(M r)^2}~,
\label{eq:O} 
\end{align}
where the prime  denotes the derivative with respect to the radial coordinate $r$. 
Equivalently, we can write the spin connection in the form 
\begin{align}
\omega_i{}^{ab}(x)=\frac{1}{M^2}\Omega(r) ~\frac12 x^jR_{ij}{}^{ab} (0)
\label{eq:O-RN}
\end{align}
where the prefactor reads
\begin{align}
\frac{1}{M^2}\Omega(r) 
= 2 \frac{1-\cos(Mr)}{(Mr)^2} &= \sum_{n=0}^\infty\frac{2(-)^n}{(2(n+1))!} (Mr)^{2n}\nonumber\\
& =1-\frac{(Mr)^2}{12}+\frac{(Mr)^4}{360}-\frac{(Mr)^6}{20160}+\cdots
\end{align}
and a power of $M^2$ is absorbed by $R_{ij}{}^{ab} (0)$.

Note that the vielbein~\eqref{eq:viel} with \eqref{eq:l-MS}
and the spin connection~\eqref{eq:O-RN} satisfy the Fock-Schwinger gauge conditions
\begin{align}
& e^a_i(x) x^i = \delta^a_i x^i\nonumber \\
& x^i \omega_i{}^{ab}(x)=0~.
\label{eq:FS}
\end{align}

\section{Wick contractions}
\label{app-B}
 We collect here the perturbative contributions up to order $\beta^8$  involved in the transition amplitude~\eqref{eq:dkernel},
 where $\la ...\ra_c$ indicates connected correlation functions. We use the
 abbreviation $\Delta(\tau_1,\tau_2)\equiv\Delta_{12}$ for the propagator,
 and $\int =\int_0^1 \!\! d\tau_1$,   $\int \!\! \int   =\int_0^1 \!\! d\tau_1\int_0^1 \!\! d\tau_2$,  
and so on, for multiple integrals. We also set $M=1$, as the dependence on $M$ is easily restored.
{\allowdisplaybreaks
\begin{align}
& S_0  = \beta k_0~,\\
&\langle S_2\rangle = -\beta^2 k_2\, d \underbrace{\int \Delta_{11}}_{-\frac{1}{6}}~, \\
& \langle S_4 \rangle =\beta^3 k_4\, d(d+2) \underbrace{\int \Delta^2_{11}}_{\frac{1}{30}} ~,\\
&\langle S_6 \rangle =- \beta^4 k_6\, {d(d+2)(d+4)}
 \underbrace{\int \Delta^3_{11}}_{-\frac{1}{140}} ~,\\
 &\frac12 \langle S_2^2 \rangle_c = \beta^4 k_2^2\, d 
 \underbrace{\int \!\! \!\!\int \Delta^2_{12}}_{\frac{1}{90}}~,\\
&\langle S_8 \rangle = \beta^5 k_8\, {d(d+2)(d+4)(d+6)}
 \underbrace{\int \Delta^4_{11}}_{\frac{1}{630}} ~,
\\&
\langle S_4 S_2 \rangle_c = - \beta^5 k_4 k_2\, 4d (d+2)
 \underbrace{ \int \!\! \!\!\int
\Delta^2_{12} \Delta_{22} }_{-\frac{1}{420}}~,\\
&\la S_{10}\ra =- \beta^6 k_{10}\, {d(d+2)(d+4)(d+6)(d+8)}
\underbrace{\int \Delta_{11}^5}_{-\frac{1}{2772}} ~,
\\&
\la S_6 S_2\ra_c = \beta^6  k_6 k_2\,  6d (d+2)(d + 4)
\underbrace{\int \!\! \!\!\int
\Delta^2_{12}\Delta^2_{22}}_{\frac{1}{1890}}~,
\\&
\frac{1}{2} \la S_4^2\ra_c = \frac{\beta^6}{2} \  k_4^2   \Biggl ( 8 d(d+2) 
\underbrace{\int \!\! \!\!\int \Delta_{12}^4}_{\frac{1}{3150}} 
+\, 8d(d+2)^2
\underbrace{\int \!\! \!\!\int \Delta_{11} \Delta_{12}^2 \Delta_{22}}_{\frac{13}{25200}} 
\Biggr)~,\\&
\frac{1}{3!} \la S_2^3\ra_c = -\frac{\beta^6}{3!}  k_2^3 \, 8 d 
\underbrace{\int \!\! \!\!\int\!\! \!\!\int
\Delta_{12}\Delta_{23}\Delta_{31}}_{-\frac{1}{945}}~,
\\&
\la S_{12}\ra = \beta^7 k_{12}\,
{d(d+2)(d+4)(d+6)(d+8)(d+10)}
\underbrace{\int \Delta_{11}^6}_{\frac{1}{12012}} ~,
\\&
\la S_8 S_2\ra_c  = -\beta^7 k_{8}k_{2}\,
{8 d(d+2)(d+4)(d+6)} 
\underbrace{\int \!\! \!\!\int \Delta_{12}^2 \Delta^3_{11}}_{-\frac{1}{8316}} ~,
\\&
\la S_6 S_4\ra_c  = -\beta^7 k_{6}k_{4} \Biggl (
{12 d (d+2)^2 (d+4)}  
\underbrace{\int \!\! \!\!\int 
 \Delta_{11}^2 \Delta_{12}^2 \Delta_{22}}_{-\frac{2}{17325}} 
\ccr &\hskip  3.7cm
+ {24d(d+2)(d+4)} 
\underbrace{
\int \!\! \!\!\int \Delta_{11} \Delta_{12}^4}_{-\frac{1}{13860}} 
\Biggr )~,
\\&
\frac12\la S_4 S_2^2 \ra_c =  \beta^7 k_4 k_2^2 \,  4 d(d+2) \Biggl (
2 \underbrace{
 \int \!\! \!\!\int \!\! \!\!\int
\,  \Delta_{12} \Delta_{13} \Delta_{23}  \Delta_{33}}_{\frac{13}{56700}} 
+ 
\underbrace{
\int \!\! \!\!\int \!\! \!\!\int
\,  \Delta^2_{13} \Delta^2_{23} }_{\frac{1}{5670}} 
\Biggr ) ~,
\\&
\la S_{14}\ra =
 -\beta^8 k_{14}\, {d(d+2)(d+4)(d+6)(d+8)(d+10)(d+12)}
\underbrace{\int  \Delta_{11}^7}_{-\frac{1}{51480}} ~,\ccr
\\&
\la S_{10} S_2\ra_c  = \beta^8 k_{10}k_{2}\,
  {10 d (d+2) (d+4)(d+6) (d+8)}  
\underbrace{
\int \!\! \!\!\int  \Delta_{11}^4 \Delta_{12}^2 }_{\frac{1}{36036}} ~,  
\\&
\la S_8 S_4\ra_c  =  
\beta^8 k_{8}k_{4}
\Biggl (  {16 d (d+2)^2 (d+4)(d+6)}  
 \underbrace{
\int \!\! \!\!\int \Delta_{11}^3 \Delta_{12}^2 \Delta_{22}}_{\frac{19}{720720}}  
\ccr
&\hskip 3.4cm +
{48d(d+2)(d+4)(d+6)} 
\underbrace{
\int \!\! \!\!\int  \Delta^2_{11} \Delta_{12}^4}_{\frac{1}{60060}} 
\Biggr )~,
\\&
\frac12 \la S_6^2 \ra_c = \frac12 \beta^8 k_6^2
\Biggl (  
{18 d (d+2)^2 (d+4)^2}  
 \underbrace{ \int \!\! \!\!\int
 \Delta_{12}^2 \Delta_{11}^2 \Delta_{22}^2}_{\frac{491}{18918900}}  
\ccr
&\quad\quad\quad\ +
{72 d (d+2) (d+4)^2}  
 \underbrace{ \int \!\! \!\!\int  \Delta_{12}^4 \Delta_{11} \Delta_{22}}_{\frac{25}{1513512}}  
+ \, 48 d (d+2) (d+4)  
 \underbrace{ \int \!\! \!\!\int  \Delta_{12}^6}_{\frac{1}{84084}}  
\Biggr )~,
\\&
\frac12 \la S_6 S^2_2\ra_c =  
 -\frac12 \beta^8 k_6 k_2^2\, 
24 d (d+2)(d+4) \Biggl(
\underbrace{ \int \!\! \!\!\int \!\! \!\!\int  \Delta^2_{11}\Delta_{12}\Delta_{13}\Delta_{23}}_{-\frac{8}{155925}}
+
\underbrace{ \int \!\! \!\!\int \!\! \!\!\int  \Delta_{11}\Delta^2_{12}\Delta^2_{13}}_{-\frac{1}{24948}}
 \Biggr)~,
\\&
\frac12 \la S_4^2 S_2\ra_c =  
 -\frac12 \beta^8 k_4^2 k_2 
\Biggl (   
32 d (d+2)^2 \Biggl (
\underbrace{ \int \!\! \!\!\int \!\! \!\!\int  \Delta^2_{12}\Delta^2_{23}\Delta_{33}}_{-\frac{2}{51975}}
+
\underbrace{ \int \!\! \!\!\int \!\! \!\!\int  \Delta_{12}\Delta_{13}\Delta_{23}\Delta_{22}\Delta_{33}}_{-\frac{83}{1663200}}
 \Biggr )
\ccr
&  \hskip 4.3cm +
64 d (d+2) 
\underbrace{ \int \!\! \!\!\int \!\! \!\!\int  \Delta_{12}\Delta_{13}\Delta^3_{23}}_{-\frac{1}{34650}}
\Biggr )~,
 \\&
\frac{1}{4!} \la S_2^4\ra_c  = \frac{1}{4!} \beta^8 k_2^4\, 48 d 
\underbrace{ \int \!\! \!\!\int \!\! \!\!\int   \!\! \!\!\int   
\Delta_{12}\Delta_{23}\Delta_{34}\Delta_{41}}_{\frac{1}{9450}} ~.
\end{align}
}
Inserting now  the values of the coupling constants $k_{2l}$  one obtains the final result in~\eqref{eq:K00}.



\begin{thebibliography}{99}

\bibitem{Bastianelli:2017wsy}
  F.~Bastianelli, O.~Corradini and E.~Vassura,
  ``Quantum mechanical path integrals in curved spaces and the type-A trace anomaly,''
  JHEP {\bf 1704} (2017) 050
  doi:10.1007/JHEP04(2017)050
  [arXiv:1702.04247 [hep-th]].
   
\bibitem{Bastianelli:2017xhi}
  F.~Bastianelli and O.~Corradini,
  ``On the simplified path integral on spheres,''
  Eur.\ Phys.\ J.\ C {\bf 77} (2017) no.11,  731
  doi:10.1140/epjc/s10052-017-5307-6
  [arXiv:1708.03557 [hep-th]].
  
\bibitem{Guven:1987en}
  J.~Guven,
  ``Calculating the effective action for a selfinteracting scalar quantum field theory in a curved background pace-time,''
  Phys.\ Rev.\ D {\bf 37} (1988) 2182.
  doi:10.1103/PhysRevD.37.2182
  
\bibitem{Copeland:1985ua}
  E.~J.~Copeland and D.~J.~Toms,
  ``The conformal anomaly in higher dimensions,''
  Class.\ Quant.\ Grav.\  {\bf 3} (1986) 431.
  doi:10.1088/0264-9381/3/3/017
  
\bibitem{Cappelli:2000fe}
  A.~Cappelli and G.~D'Appollonio,
  ``On the trace anomaly as a measure of degrees of freedom,''
  Phys.\ Lett.\ B {\bf 487} (2000) 87
  doi:10.1016/S0370-2693(00)00809-1
  [hep-th/0005115].
  
\bibitem{Aros:2011iz}
  R.~Aros and D.~E.~Diaz,
  ``Determinant and Weyl anomaly of Dirac operator: a holographic derivation,''
  J.\ Phys.\ A {\bf 45} (2012) 125401
  doi:10.1088/1751-8113/45/12/125401
  [arXiv:1111.1463 [math-ph]].
     
\bibitem{Diaz:2008hy}
  D.~E.~Diaz,
  ``Polyakov formulas for GJMS operators from AdS/CFT,''
  JHEP {\bf 0807} (2008) 103
  doi:10.1088/1126-6708/2008/07/103
  [arXiv:0803.0571 [hep-th]].
  
\bibitem{Dowker:2010qy}
  J.~S.~Dowker,
  ``Determinants and conformal anomalies of GJMS operators on spheres,''
  J.\ Phys.\ A {\bf 44} (2011) 115402
  doi:10.1088/1751-8113/44/11/115402
  [arXiv:1010.0566 [hep-th]].

\bibitem{Schubert:2001he}
  C.~Schubert,
  ``Perturbative quantum field theory in the string inspired formalism,''
  Phys.\ Rept.\  {\bf 355} (2001) 73
  doi:10.1016/S0370-1573(01)00013-8
  [hep-th/0101036].

\bibitem{Dunne:2005sx}
  G.~V.~Dunne and C.~Schubert,
  ``Worldline instantons and pair production in inhomogeneous fields,''
  Phys.\ Rev.\ D {\bf 72} (2005) 105004
  doi:10.1103/PhysRevD.72.105004
  [hep-th/0507174].
  
\bibitem{Dunne:2006st}
  G.~V.~Dunne, Q.~h.~Wang, H.~Gies and C.~Schubert,
  ``Worldline instantons. II. The Fluctuation prefactor,''
  Phys.\ Rev.\ D {\bf 73} (2006) 065028
  doi:10.1103/PhysRevD.73.065028
  [hep-th/0602176].
  
\bibitem{Ahmadiniaz:2012xp}
  N.~Ahmadiniaz and C.~Schubert,
  ``A covariant representation of the Ball-Chiu vertex,''
  Nucl.\ Phys.\ B {\bf 869} (2013) 417
  doi:10.1016/j.nuclphysb.2012.12.019
  [arXiv:1210.2331 [hep-ph]].
  
\bibitem{Bastianelli:2013pta}
  F.~Bastianelli, R.~Bonezzi, O.~Corradini and E.~Latini,
  ``Particles with non abelian charges,''
  JHEP {\bf 1310} (2013) 098
  doi:10.1007/JHEP10(2013)098
  [arXiv:1309.1608 [hep-th]].
  
\bibitem{Ahmadiniaz:2015xoa}
  N.~Ahmadiniaz, F.~Bastianelli and O.~Corradini,
  ``Dressed scalar propagator in a non-Abelian background from the worldline formalism,''
  Phys.\ Rev.\ D {\bf 93} (2016) no.2,  025035
   Addendum: [Phys.\ Rev.\ D {\bf 93} (2016) no.4,  049904]
  doi:10.1103/PhysRevD.93.049904, 10.1103/PhysRevD.93.025035
  [arXiv:1508.05144 [hep-th]].
  
\bibitem{Ahmadiniaz:2015kfq}
  N.~Ahmadiniaz, A.~Bashir and C.~Schubert,
  ``Multiphoton amplitudes and generalized Landau-Khalatnikov-Fradkin transformation in scalar QED,''
  Phys.\ Rev.\ D {\bf 93} (2016) no.4,  045023
  doi:10.1103/PhysRevD.93.045023
  [arXiv:1511.05087 [hep-ph]].
  
\bibitem{Edwards:2017bte}
  J.~P.~Edwards and C.~Schubert,
  ``One-particle reducible contribution to the one-loop scalar propagator in a constant field,''
  Nucl.\ Phys.\ B {\bf 923} (2017) 339
  doi:10.1016/j.nuclphysb.2017.08.002
  [arXiv:1704.00482 [hep-th]].
  
\bibitem{Ahmadiniaz:2017rrk}
  N.~Ahmadiniaz, F.~Bastianelli, O.~Corradini, J.~P.~Edwards and C.~Schubert,
  ``One-particle reducible contribution to the one-loop spinor propagator in a constant field,''
  Nucl.\ Phys.\ B {\bf 924} (2017) 377
  doi:10.1016/j.nuclphysb.2017.09.012
  [arXiv:1704.05040 [hep-th]].
 
\bibitem{Bastianelli:2002fv}
  F.~Bastianelli and A.~Zirotti,
  ``Worldline formalism in a gravitational background,''
  Nucl.\ Phys.\ B {\bf 642} (2002) 372
  doi:10.1016/S0550-3213(02)00683-1
  [hep-th/0205182].

\bibitem{Bastianelli:2002qw}
  F.~Bastianelli, O.~Corradini and A.~Zirotti,
  ``Dimensional regularization for N=1 supersymmetric sigma models and the worldline formalism,''
  Phys.\ Rev.\ D {\bf 67} (2003) 104009
  doi:10.1103/PhysRevD.67.104009
  [hep-th/0211134].

\bibitem{Bastianelli:2005vk}
  F.~Bastianelli, P.~Benincasa and S.~Giombi,
  ``Worldline approach to vector and antisymmetric tensor fields,''
  JHEP {\bf 0504} (2005) 010
  doi:10.1088/1126-6708/2005/04/010
  [hep-th/0503155].

\bibitem{Bastianelli:2005uy}
  F.~Bastianelli, P.~Benincasa and S.~Giombi,
  ``Worldline approach to vector and antisymmetric tensor fields. II.,''
  JHEP {\bf 0510} (2005) 114
  doi:10.1088/1126-6708/2005/10/114
  [hep-th/0510010].

\bibitem{Bastianelli:2004zp}
  F.~Bastianelli and C.~Schubert,
  ``One loop photon-graviton mixing in an electromagnetic field: Part 1,''
  JHEP {\bf 0502} (2005) 069
  doi:10.1088/1126-6708/2005/02/069
  [gr-qc/0412095].

\bibitem{Hollowood:2007ku} 
  T.~J.~Hollowood and G.~M.~Shore,
  ``The refractive index of curved spacetime: the fate of causality in QED,''
  Nucl.\ Phys.\ B {\bf 795}, 138 (2008)
  doi:10.1016/j.nuclphysb.2007.11.034
  [arXiv:0707.2303 [hep-th]].

\bibitem{Bastianelli:2008cu}
  F.~Bastianelli, J.~M.~Davila and C.~Schubert,
  ``Gravitational corrections to the Euler-Heisenberg Lagrangian,''
  JHEP {\bf 0903} (2009) 086
  doi:10.1088/1126-6708/2009/03/086
  [arXiv:0812.4849 [hep-th]].

\bibitem{Bastianelli:2013tsa}
  F.~Bastianelli and R.~Bonezzi,
  ``One-loop quantum gravity from a worldline viewpoint,''
  JHEP {\bf 1307} (2013) 016
  doi:10.1007/JHEP07(2013)016
  [arXiv:1304.7135 [hep-th]].
  
\bibitem{Bastianelli:2007pv}
  F.~Bastianelli, O.~Corradini and E.~Latini,
  ``Higher spin fields from a worldline perspective,''
  JHEP {\bf 0702} (2007) 072
  doi:10.1088/1126-6708/2007/02/072
  [hep-th/0701055].
  
\bibitem{Bastianelli:2008nm}
  F.~Bastianelli, O.~Corradini and E.~Latini,
  ``Spinning particles and higher spin fields on (A)dS backgrounds,''
  JHEP {\bf 0811} (2008) 054
  doi:10.1088/1126-6708/2008/11/054
  [arXiv:0810.0188 [hep-th]].
  
\bibitem{Corradini:2010ia}
  O.~Corradini,
  ``Half-integer higher spin fields in (A)dS from spinning particle models,''
  JHEP {\bf 1009} (2010) 113
  doi:10.1007/JHEP09(2010)113
  [arXiv:1006.4452 [hep-th]].
   
\bibitem{Bastianelli:2012bn}
  F.~Bastianelli, R.~Bonezzi, O.~Corradini and E.~Latini,
  ``Effective action for higher spin fields on (A)dS backgrounds,''
  JHEP {\bf 1212} (2012) 113
  doi:10.1007/JHEP12(2012)113
  [arXiv:1210.4649 [hep-th]].
  
\bibitem{Bonezzi:2017mwr}
  R.~Bonezzi,
  ``Induced action for conformal higher spins from worldline path integrals,''
  Universe {\bf 3} (2017) no.3,  64
  doi:10.3390/universe3030064
  [arXiv:1709.00850 [hep-th]].
    
\bibitem{Bonora:2018uwx} 
  L.~Bonora, M.~Cvitan, P.~Dominis Prester, S.~Giaccari, M.~Paulisic and T.~Stemberga,
  ``Worldline quantization of field theory, effective actions and $L_\infty$ structure,''
  arXiv:1802.02968 [hep-th].
  
\bibitem{Bastianelli:2006rx}
  F.~Bastianelli and P.~van Nieuwenhuizen,
  ``Path integrals and anomalies in curved space,''
   Cambridge University Press, Cambridge, UK (2006).
  
\bibitem{Bastianelli:2001tb}
  F.~Bastianelli and N.~D.~Hari Dass,
  ``Simplified method for trace anomaly calculations in $d \leq 6$,''
  Phys.\ Rev.\ D {\bf 64} (2001) 047701
  doi:10.1103/PhysRevD.64.047701
  [hep-th/0104234].

\bibitem{Berezin:1976eg}
  F.~A.~Berezin and M.~S.~Marinov,
  ``Particle spin dynamics as the Grassmann variant of classical mechanics,''
  Annals Phys.\  {\bf 104} (1977) 336.
  doi:10.1016/0003-4916(77)90335-9
  
\bibitem{Barducci:1976qu}
  A.~Barducci, R.~Casalbuoni and L.~Lusanna,
  ``Supersymmetries and the pseudoclassical relativistic electron,''
  Nuovo Cim.\ A {\bf 35} (1976) 377.
  doi:10.1007/BF02730291
  
\bibitem{Brink:1976sz}
  L.~Brink, S.~Deser, B.~Zumino, P.~Di Vecchia and P.~S.~Howe,
  ``Local supersymmetry for spinning particles,''
  Phys.\ Lett.\  {\bf 64B} (1976) 435
   Erratum: [Phys.\ Lett.\  {\bf 68B} (1977) 488].
  doi:10.1016/0370-2693(76)90115-5
  
\bibitem{Deser:1993yx}
  S.~Deser and A.~Schwimmer,
  ``Geometric classification of conformal anomalies in arbitrary dimensions,''
  Phys.\ Lett.\ B {\bf 309} (1993) 279
  doi:10.1016/0370-2693(93)90934-A
  [hep-th/9302047].
  
\bibitem{Boulanger:2007ab}
  N.~Boulanger,
  ``Algebraic Classification of Weyl Anomalies in Arbitrary Dimensions,''
  Phys.\ Rev.\ Lett.\  {\bf 98} (2007) 261302
  doi:10.1103/PhysRevLett.98.261302
  [arXiv:0706.0340 [hep-th]].
  
\bibitem{Boulanger:2007st}
  N.~Boulanger,
  ``General solutions of the Wess-Zumino consistency condition for the Weyl anomalies,''
  JHEP {\bf 0707} (2007) 069
  doi:10.1088/1126-6708/2007/07/069
  [arXiv:0704.2472 [hep-th]].
    
\bibitem{Fujikawa:1980vr}
  K.~Fujikawa,
  ``Comment on chiral and conformal anomalies,''
  Phys.\ Rev.\ Lett.\  {\bf 44} (1980) 1733.
  doi:10.1103/PhysRevLett.44.1733
    
\bibitem{Bastianelli:1991be}
  F.~Bastianelli,
  ``The path integral for a particle in curved spaces and Weyl anomalies,''
  Nucl.\ Phys.\ B {\bf 376} (1992) 113
  doi:10.1016/0550-3213(92)90070-R
  [hep-th/9112035].
  
\bibitem{Bastianelli:1992ct}
  F.~Bastianelli and P.~van Nieuwenhuizen,
  ``Trace anomalies from quantum mechanics,''
  Nucl.\ Phys.\ B {\bf 389} (1993) 53
  doi:10.1016/0550-3213(93)90285-W
  [hep-th/9208059].
  
\bibitem{Diaz:1989nx}
  A.~Diaz, W.~Troost, P.~van Nieuwenhuizen and A.~Van Proeyen,
  ``Understanding Fujikawa regulators from Pauli-Villars regularization of ghost loops,''
  Int.\ J.\ Mod.\ Phys.\ A {\bf 4} (1989) 3959.
  doi:10.1142/S0217751X8900162X
  
\bibitem{Zamolodchikov:1986gt}
  A.~B.~Zamolodchikov,
  ``Irreversibility of the flux of the renormalization group in a 2D field theory,''
  JETP Lett.\  {\bf 43} (1986) 730
   [Pisma Zh.\ Eksp.\ Teor.\ Fiz.\  {\bf 43} (1986) 565].
  
\bibitem{Komargodski:2011vj}
  Z.~Komargodski and A.~Schwimmer,
  ``On renormalization group flows in four dimensions,''
  JHEP {\bf 1112} (2011) 099
  doi:10.1007/JHEP12(2011)099
  [arXiv:1107.3987 [hep-th]].

\bibitem{Camporesi:1992tm}
  R.~Camporesi,
  ``The spinor heat kernel in maximally symmetric spaces,''
  Commun.\ Math.\ Phys.\  {\bf 148} (1992) 283.
  doi:10.1007/BF02100862
   
\bibitem{Howe:1989vn}
  P.~S.~Howe, S.~Penati, M.~Pernici and P.~K.~Townsend,
  ``A particle mechanics description of antisymmetric tensor fields,''
  Class.\ Quant.\ Grav.\  {\bf 6} (1989) 1125.
  doi:10.1088/0264-9381/6/8/012

\bibitem{Gershun:1979fb}
  V.~D.~Gershun and V.~I.~Tkach,
  ``Classical and quantum dynamics of particles with arbitrary spin,''
  JETP Lett.\  {\bf 29} (1979) 288
   [Pisma Zh.\ Eksp.\ Teor.\ Fiz.\  {\bf 29} (1979) 320].
  
\bibitem{Howe:1988ft}
  P.~S.~Howe, S.~Penati, M.~Pernici and P.~K.~Townsend,
  ``Wave equations for arbitrary spin from quantization of the extended supersymmetric spinning particle,''
  Phys.\ Lett.\ B {\bf 215} (1988) 555.
  doi:10.1016/0370-2693(88)91358-5
  
\bibitem{Siegel:1988ru}
  W.~Siegel,
  ``Conformal invariance of extended spinning particle mechanics,''
  Int.\ J.\ Mod.\ Phys.\ A {\bf 3} (1988) 2713.
  doi:10.1142/S0217751X88001132
    
\bibitem{Siegel:1988gd} 
  W.~Siegel,
  ``All free conformal representations in all dimensions,''
  Int.\ J.\ Mod.\ Phys.\ A {\bf 4}, 2015 (1989).
  doi:10.1142/S0217751X89000819
 
\bibitem{Kuzenko:1995mg}
  S.~M.~Kuzenko and Z.~V.~Yarevskaya,
  ``Conformal invariance, N extended supersymmetry and massless spinning particles in anti-de Sitter space,''
  Mod.\ Phys.\ Lett.\ A {\bf 11} (1996) 1653
  doi:10.1142/S0217732396001648
  [hep-th/9512115].

\bibitem{Bastianelli:2011cc}
  F.~Bastianelli, R.~Bonezzi, O.~Corradini and E.~Latini,
  ``Extended SUSY quantum mechanics: transition amplitudes and path integrals,''
  JHEP {\bf 1106} (2011) 023
  doi:10.1007/JHEP06(2011)023
  [arXiv:1103.3993 [hep-th]].
    
\bibitem{Bastianelli:2006hq}
  F.~Bastianelli, O.~Corradini and P.~A.~G.~Pisani,
  ``Worldline approach to quantum field theories on flat manifolds with boundaries,''
  JHEP {\bf 0702} (2007) 059
  doi:10.1088/1126-6708/2007/02/059
  [hep-th/0612236].
    
\bibitem{Bastianelli:2008vh}
  F.~Bastianelli, O.~Corradini, P.~A.~G.~Pisani and C.~Schubert,
  ``Scalar heat kernel with boundary in the worldline formalism,''
  JHEP {\bf 0810} (2008) 095
  doi:10.1088/1126-6708/2008/10/095
  [arXiv:0809.0652 [hep-th]].
 
\bibitem{Solodukhin:2015eca}
  S.~N.~Solodukhin,
  ``Boundary terms of conformal anomaly,''
  Phys.\ Lett.\ B {\bf 752} (2016) 131
  doi:10.1016/j.physletb.2015.11.036
  [arXiv:1510.04566 [hep-th]].
  
\bibitem{Fursaev:2016inw} 
  D.~V.~Fursaev and S.~N.~Solodukhin,
  ``Anomalies, entropy and boundaries,''
  Phys.\ Rev.\ D {\bf 93}, no. 8, 084021 (2016)
  doi:10.1103/PhysRevD.93.084021
  [arXiv:1601.06418 [hep-th]].
  
\bibitem{Rodriguez-Gomez:2017kxf} 
  D.~Rodriguez-Gomez and J.~G.~Russo,
  ``Free energy and boundary anomalies on $\mathbb{S}^a\times \mathbb{H}^b$ spaces,''
  JHEP {\bf 1710}, 084 (2017)
  doi:10.1007/JHEP10(2017)084
  [arXiv:1708.00305 [hep-th]].
  
\bibitem{Rodriguez-Gomez:2017aca}
  D.~Rodriguez-Gomez and J.~G.~Russo,
  ``Boundary conformal anomalies on hyperbolic spaces and euclidean balls,''
  JHEP {\bf 1712} (2017) 066
  doi:10.1007/JHEP12(2017)066
  [arXiv:1710.09327 [hep-th]].
  
  \bibitem{Eisenhart:1965}
 L. P. Eisenhart, ``Riemannian geometry," Princeton University Press,  Princeton USA (1965).
 
 \bibitem{Petrov:1969}
A. Z. Petrov, ``Einstein spaces", Pergamon Press, Oxford UK (1969).

\bibitem{Muller:1997zk}
  U.~Muller, C.~Schubert and A.~M.~E.~van de Ven,
  ``A closed formula for the Riemann normal coordinate expansion,''
  Gen.\ Rel.\ Grav.\  {\bf 31} (1999) 1759
  doi:10.1023/A:1026718301634
  [gr-qc/9712092].

\bibitem{AlvarezGaume:1981hn}
  L.~Alvarez-Gaume, D.~Z.~Freedman and S.~Mukhi,
  ``The background field method and the ultraviolet structure of the supersymmetric nonlinear sigma model,''
  Annals Phys.\  {\bf 134} (1981) 85.
  doi:10.1016/0003-4916(81)90006-3
  
\bibitem{Howe:1986vm}
  P.~S.~Howe, G.~Papadopoulos and K.~S.~Stelle,
  ``The background field method and the nonlinear $\sigma$ model,''
  Nucl.\ Phys.\ B {\bf 296} (1988) 26.
  doi:10.1016/0550-3213(88)90379-3
    
\bibitem{Bastianelli:2000dw}
  F.~Bastianelli and O.~Corradini,
  ``6-D trace anomalies from quantum mechanical path integrals,''
  Phys.\ Rev.\ D {\bf 63} (2001) 065005
  doi:10.1103/PhysRevD.63.065005
  [hep-th/0010118].
      
\end{thebibliography}
\end{document}